\documentclass[twocolumn,showpacs,preprintnumbers,amsmath,amssymb,APSl,prd,nofootinbib,superscriptaddress]{revtex4-1}
\pdfoutput=1
\usepackage{epsfig}
\usepackage{epstopdf}
\usepackage{hyperref}
\usepackage{amssymb}
\usepackage{amsmath}
\usepackage{amsfonts}
\usepackage{graphicx}
\usepackage{bm}
\usepackage[usenames,dvipsnames]{color}
\usepackage{hyperref}
\usepackage{ifpdf}
\def	\be	{\begin{equation}}
\def	\ee	{\end{equation}}
\def	\bqt	{\begin{quote}}
\def	\eqt	{\end{quote}}

\begin{document}

\title{$\kappa(R,T)$ gravity}

\author{Gin\'{e}s R.P\'{e}rez Teruel}

\affiliation{Departamento de Matem\'{a}ticas, IES Can\'{o}nigo Manch\'{o}n, Crevillente-03330, Alicante, Spain} 
\email{gines.landau@gmail.com}
\begin{abstract}
\begin{center}
{\bf Abstract}
\end{center}
\noindent
In this note we explore a modified theory of gravitation that is not based on the least action principle, but on a natural generalization of the original Einstein's field equations. This approach leads to the non-covariant conservation of the stress-energy tensor, a feature shared with other Lagrangian theories of gravity such as the $f(R,T)$ case. We consider the cosmological implications of a pair of particular models within this theory, and we show that they have some interesting properties. In particular, for some of the studied models we find that the density is bounded from above, and cannot exceed a maximum value that depends on certain physical constants. In the last part of the work we compare the theory to the $f(R,T)$ case and show that they lead to different predictions for the motion of test particles.
\end{abstract}

\maketitle
\section{Introduction}
\label{Introduction} 
\thispagestyle{empty}

\noindent
The least action principle has become one of the most powerful tools to build a physical theory and also their possible generalizations. Among the numerous advantages of the Lagrangian formalism we list the direct implementation of symmetries and the derivation of general conservation laws. Nevertheless, there is no reason to believe that ordinary symmetries and/or standard conservation laws will always hold in a final theory of Nature. In this sense, it is interesting to recall that Einstein did not originally followed a variational principle in the derivation of General Relativity (GR)\cite{Renn,Sauer,Einstein}. Instead, he arrived to the correct field equations following a very different approach, one that succeeded with the eventual addition of a trace term directly in the field equations. Indeed, it is well known that the equivalence principle and general covariance were the foundational concepts of the theory, and the variational principle, i.e, the Einstein-Hilbert Action (EHA), was discovered and incorporated to the theory when the correct field equations had already been derived.\cite{Hilbert} 
Similarly, the other classical field theory, namely, Maxwell's Eletrodynamics (ME), was only completed after the addition of a source term (the Maxwell displacement current), and it wasn't either originally conceived from any variational principle.\cite{Maxwell}\\  
In light of these historical facts, there is no reason to reject the search for an alternative approach,one different from the Lagrangian formalism. In this sense, a more general theory could be formulated following an alternative but consistent line of reasoning, and the variational principle (or something alike) could be incorporated in the last stages of the completion of the theory in order to strengthen and to reinforce the formalism. The fact that both the GR and ME field equations were found without resorting to a variational principle sends us the message that maybe a different approach deserves to being taken into account.\\
Regarding the case of GR and their extensions, there exists a high degree of arbitrariness in the choice of the specific generalized gravity Lagrangian. One of the most natural strategies is to replace the curvature scalar $R$, by a function $f(R)$ in the action\cite{Buchdahl,DeFelice,Capo,Olmo,Lobo,Sotiriou,Mafia}. This approach leads to field equations that give rise to a rich phenomenology (both in the metric and metric-affine or Palatini formalisms), although some of the beauty and simplicity of the original theory are lost in the process. Regardless the specific choice among the innumerable possible generalized Lagrangians, the resulting field equations are much more complicated than those of GR.\\
In view of the above, given the arbitrariness in the choice of the possible gravity Lagrangian, we do not begin from a Lagrangian formalism to look for a modification of GR. Rather, we follow Maxwell's and Einstein's original approaches of adding new possible source terms directly in the field equations, this means that we focus on a similar strategy that proved successful in the completion of the classical Electromagnetic theory, and in the first correct derivation of GR. In particular, the theory considered here is based on a quite natural extension of GR, where the modification of the field equations involves the addition of terms that only include the curvature scalar and the trace of the stress-energy tensor. Therefore, in vacuum the field equations boil down to those of GR, but in presence of matter there can be significant departures, in particular the stress-energy tensor will not be covariantly conserved in the general case. There are several examples in the literature of gravitational theories of this kind. An example is Rastall's gravitational theory\cite{Rastall,Fabris}, which is also non-conservative since the divergence of $T_{\mu\nu}$ does not vanish in general. Another more recent example is given by the so-called $f(R,T)$ modified theories of gravity\cite{Harko,Sharif,Shabani,Shabani2,Shabani3}. We should mention that this note presents an study of a modified theory of gravity at a preliminary level, it is a first step in a different direction than that adopted by the usual modified gravity theories, and further investigations are required to discuss in more detail some important aspects that are addressed in this work. It is becoming increasingly clear that quantum field theories without a traditional Lagrangian description are important and even populate much of the QFT landscape. They also offer new opportunities in the search of new type of 4-manifold invariants\cite{Gukov,Razamat}. This fact represents another good motivation to explore here an example of a Non-Lagrangian modified gravitational theory.
\section{Definitions and Field Equations}
Our framework is based on the following field equations
\begin{equation}\label{FieldEquations}
R_{\mu\nu}-\frac{1}{2}Rg_{\mu\nu}-\Lambda g_{\mu\nu}=\kappa(R,T)T_{\mu\nu}
\end{equation}
Where $R_{\mu\nu}$ is the Ricci tensor, $g_{\mu\nu}$ is the space-time metric, $\Lambda$ is a cosmological constant, $T_{\mu\nu}$ the stress-energy tensor of the matter sources, and $\kappa(R,T)$ corresponds to the Einstein gravitational constant that we are promoting to the status of a function of the traces $T\equiv g_{\mu\nu}T^{\mu\nu}$, and $R\equiv g_{\mu\nu}R^{\mu\nu}$. The possible dependence of the gravitational constant $\kappa$ on scalars means that we explore the possibility of a running gravitational constant, i.e. we generalize the original Einstein's gravitational constant, but not at the level of an action functional. A varying gravitational constant in the action leads to a Brans-Dicke type theory\cite{Brans,Brans2,Moffat}, with quite different field equations from (\ref{FieldEquations}). The field equations (\ref{FieldEquations}) imply the non-covariant conservation of $T_{\mu\nu}$. Indeed, since the left hand side of these equations is divergence-free, we have
\begin{equation}\label{nonconservation1}
\nabla{^\nu}\Big(\kappa(R,T)T_{\mu\nu}\Big)=0
\end{equation}
Then, the non-conservation of the $T_{\mu\nu}$ can be expressed as
\begin{equation}\label{nonconservation2}
\nabla^{\nu}T_{\mu\nu}=-\frac{\nabla^{\nu} \kappa(R,T)}{\kappa(R,T)}T_{\mu\nu}
\end{equation}
In what follows, some cosmological implications (homogeneous and isotropic universe for a perfect fluid) of two particular cases are analyzed.  The first model considered arises by setting, $\kappa(T)=8\pi G-\lambda T$, and corresponds to a matter-matter coupling. The second model that will be studied is characterized by a gravitational ``constant" that varies as $\kappa^{\prime}(R)=8\pi G+\alpha R$, which will provide a coupling between matter and curvature terms. We assume that the coupling constants $\lambda$, $\alpha$ are sufficiently small to be consistent with a small violation of the covariant conservation of the stress-energy tensor. Obviously, in the limit $\lambda,\alpha\rightarrow 0$, Einstein's GR is recovered.
\subsection{Modified Friedmann Equations for a general $\kappa(R,T)$ model}
If we consider an homogeneous and isotropic universe filled by a perfect fluid as the matter source, the stress-energy tensor will be 

\begin{equation}\label{PerfectFluid}
T_{\mu\nu}=(p+\rho)u_{\mu}u_{\nu}-pg_{\mu\nu}
\end{equation}
where $p$, $\rho$, and $u_{\mu}$ are the pressure, the density and the macroscopic speed of the medium, respectively. On the other hand, the standard FLRW metric for modeling the assumed properties of such a universe leads to the line element

\begin{equation}\label{FLRW}
ds^{2}=dt^{2}-a(t)^{2}\Big(\frac{1}{1-\frac{r^{2}}{K^{2}}}dr^{2}+r^{2}d\theta^{2}+r^{2}\sin^{2}\theta d\varphi^{2}\Big)
\end{equation}
\\
With these ingredients, the two independent Modified Friedmann Equations (MFE) for a general $\kappa(R,T)$ model are
\begin{equation}
\Big(\frac{\dot{a}(t)}{a(t)}\Big)^{2}+\frac{1}{K^{2}a^{2}}-\frac{\Lambda}{3}=\frac{\kappa(R,T)}{3}\rho(t)
\end{equation}
\begin{equation}
\frac{\ddot{a}(t)}{a(t)}=\frac{\Lambda}{3}-\frac{\kappa(R,T)}{6}\Big(3p(t)+\rho(t)\Big)
\end{equation}
Two independent models of the form $\kappa(T)$ and $\kappa(R)$ are analyzed in the next subsections.
\subsection{Matter-matter coupling}
Here we analyze the cosmological implications of the model, $\kappa(T)=k-\lambda T$, where $k\equiv 8\pi G$ ($c=1$) and $\lambda$ is a constant with the appropriate units. The reasons to choose a negative sign will be understood later on. Therefore, with such a choice for $\kappa(R,T)$ the field equations \ref{FieldEquations} acquire the form

\begin{equation}\label{FieldEquations2}
R_{\mu\nu}-\frac{1}{2}Rg_{\mu\nu}-\Lambda g_{\mu\nu}=\Big(8\pi G-\lambda T\Big)T_{\mu\nu}
\end{equation}

\begin{figure}[h!]
\includegraphics[width=8cm]{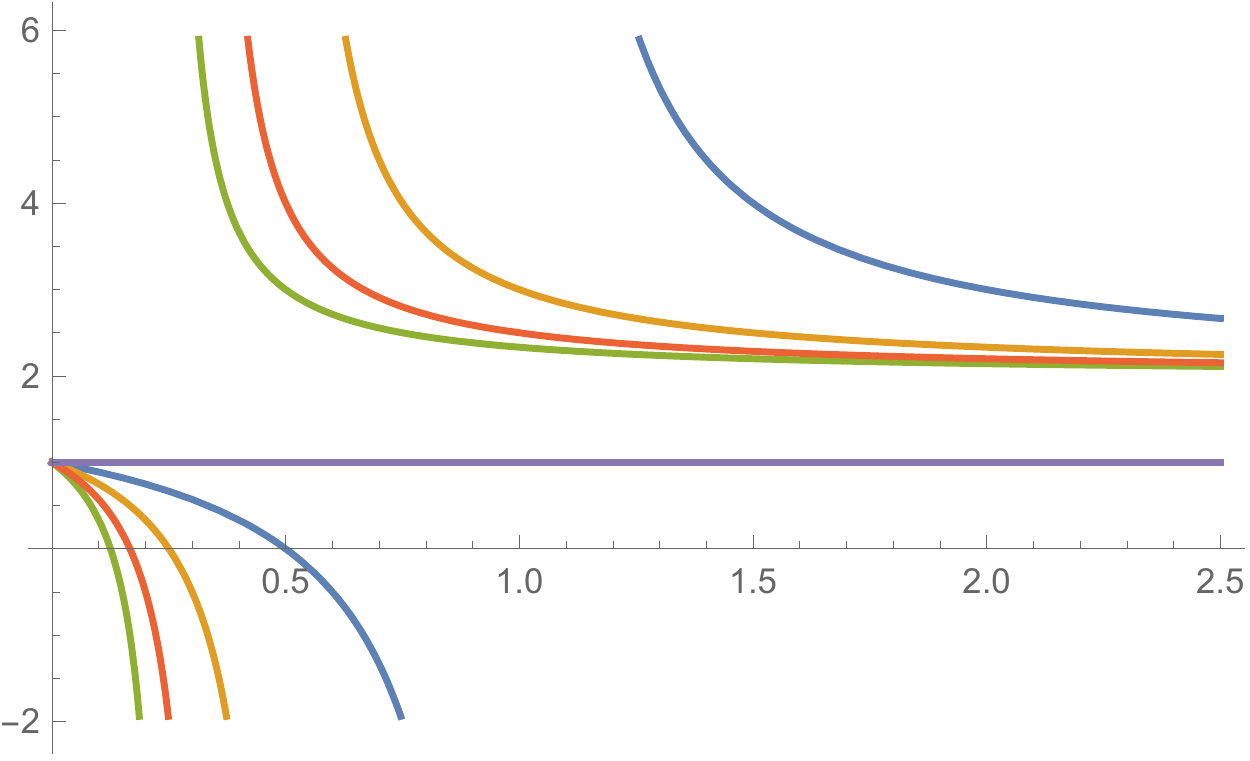}
\caption{\small{Variation of $F(\xi)$ as function of the dimensionless variable $\xi\equiv \rho/\rho_{m}$ for different values of the parameter $w$. The horizontal line corresponds to $w=1/3$. At low densities $F(\xi)\simeq 1$ and the RG behavior is recovered}}
\label{fig:a3}
\end{figure}

The MFE that arise by solving the field equations are
\begin{equation}
H^{2}=\frac{8\pi G}{3}\rho+\frac{\Lambda}{3}-\frac{1}{K^{2}a^{2}}-\frac{\lambda\rho}{3}\Big(\rho-3p\Big)
\end{equation}
\begin{equation}
\frac{\ddot{a}}{a}=-\frac{4\pi G}{3}\Big(3p+\rho\Big)+\frac{\Lambda}{3}+\frac{\lambda}{6}\Big(\rho-3p\Big)\Big(\rho+3p\Big)
\end{equation}
Where, $H=\dot{a}/{a}$. In order to better understand the physical meaning of the MFE, and for the sake of simplicity it is convenient to make use of a equation of state of the type, $p=w\rho$. Then, for the flat case ($K^{-1}=0$), we can write
\begin{equation}\label{H parameter}
H^{2}=\frac{8\pi G}{3}\rho+\frac{\Lambda}{3}-\frac{\lambda\rho^{2}}{3}\Big(1-3w\Big)
\end{equation}
\begin{equation}\label{AccelerationEquation}
\frac{\ddot{a}}{a}=-\frac{4\pi G}{3}\rho\Big(3w+1\Big)+\frac{\Lambda}{3}+\frac{\lambda}{6}\rho^{2}\Big(1+3w\Big)\Big(1-3w\Big)
\end{equation}
It is therefore clear that the new additional term introduced vanishes for $w=1/3$ (radiation-dominated universe). For $w=-1/3$ it vanishes for the second equation as well. For a radiation-dominated universe, the traceless of the stress-energy tensor implies that the MFE collapse to the standard solution of GR with a cosmological constant. On the other hand, the contribution of the new term is proportional to $\rho^{2}$, for $-1/3< w < 1/3$ this quadratic term will be positive in the acceleration equation, and this means that at sufficiently high densities it could contribute to the cosmic speed-up. However, to explain the late-time acceleration (low densities) we unavoidably need the inclusion of a cosmological constant.\\
The non-conservation of the stress-energy tensor implies a modification of the relativistic fluid equation characterized by 
\begin{equation}\label{Fluid}
\dot{\rho}+3\frac{\dot{a}}{a}\rho\Big(1+w\Big)F(\rho)=0
\end{equation}
The correction with respect to the GR case is represented by the presence of a certain function $F(\rho)$ which is explicitly given by
\begin{equation}\label{Correction}
F(\rho)=\frac{1-(1-3w)\frac{2\rho}{\rho_{m}}}{1-(1-3w)\frac{\rho}{\rho_{m}}}
\end{equation}
Where we have denoted a certain constant $\rho_{m}$ with units of density as
\begin{equation}
\rho_{m}=\frac{8\pi G}{\lambda}
\end{equation}
In Fig.\ref{fig:a3}, We plot the behavior of the correction factor $F$ as function of $\rho/\rho_{m}$. Notice that when $\rho<<\rho_{m}$, then, $F(\rho)\simeq 1$ and we essentially recover the same fluid equation of GR. However, for ultra high-densities, namely, $\rho>>\rho_{m}$, $F(\rho)\simeq 2$. For such a regime, the relativistic fluid equation becomes
\begin{equation}
\dot{\rho}+6\frac{\dot{a}}{a}\rho\Big(1+w\Big)=0
\end{equation}
This implies that, $\rho(t)\sim a(t)^{-6(1+w)}$ which suggests a much more rapid decrease of the density with the scale factor. However, the density cannot be arbitrarily large, and in general it will be of order $\rho_{m}$. This is due to the requirement $H^{2}\geq 0$ which imposes an upper bound for the density. Indeed, suppose that at sufficiently high densities we can neglect the contribution of the cosmological constant term compared to the other two in the first MFE. Then, reality of $H^{2}$ requires
\begin{equation}\label{bounds}
\frac{8\pi G}{3}\rho\geq \frac{\lambda \rho^{2}}{3}(1-3w)
\end{equation}
which implies
\begin{equation}\label{maximum density}
\rho_{\max}\sim\frac{8\pi G}{\lambda(1-3w)}=\frac{\rho_{m}}{(1-3w)}
\end{equation}
For $w\neq 1/3$. Notice that for $\rho=\rho_{\max}$, $H\sim \sqrt{\Lambda}$ and therefore the Hubble parameter behaves such as de Sitter vacuum solution. Finally, the choice of the negative sign in the function $\kappa(T)=8\pi G-\lambda T$ is due to very good reasons that turn out to be evident now. If we had chosen the positive sign, we would have that at high densities $H^{2}\sim \rho^{2}$, which is worst in terms of divergences than the GR case.\cite{Olmo2}\\
Field equations similar to (\ref{FieldEquations}), or (\ref{H parameter}-\ref{AccelerationEquation}) in cosmology, could arise from Ricci and specially from generalized Ricci-Gauss-Bonnet holographic dark energy models, see for example Refs.\cite{Gao,Nojiri2,Saridakis}
\subsubsection{\textbf{Dust solution}}
Let us solve the equations for the simplest cosmological model, which is obtained by assuming a dust universe, where $p=0$. The MFE for such a universe become
\begin{equation}
H^{2}=\frac{8\pi G}{3}\rho+\frac{\Lambda}{3}-\frac{\lambda}{3}\rho^{2}
\end{equation}
\begin{equation}
\frac{\ddot{a}}{a}=-\frac{4\pi G}{3}\rho+\frac{\Lambda}{3}+\frac{\lambda}{6}\rho^{2}
\end{equation}
Using that $\ddot{a}/a=\dot H+H^{2}$ and combining both equations we obtain
\begin{equation}
\dot{H}+\frac{3}{2}H^{2}=\frac{\Lambda}{2}
\end{equation}
These results mean that $H(t)$ for dust evolves as in the GR case, and does not depend on the value of $\lambda$. For a matter-dominated universe where we can neglect the contribution of the cosmological constant, we have $H(t)=2/3t$, $a(t)=t^{2/3}$. Then, the dust solution shows that the $\lambda$ parameter for this particular model has no effect in the expansion rate of the universe, which turns out to be identical to the GR case, and the main difference is the existence of a bound for the density, given by $\rho_{\max}=8\pi G/3\lambda$. On the other hand, the dust solution for the $f(R,T)$ theory was studied in \cite{Harko}, where for a function of the form $f(T)=\lambda T$ the Hubble parameter was found to be dependent on $\lambda$ as
\begin{equation}
H(t)_{f(R,T)}=\frac{2(8\pi+3\lambda)}{3(8\pi+2\lambda)}\frac{1}{t}
\end{equation}
Therefore, $H(t)_f>H(t)_\kappa$, i.e, the Hubble parameter in the $f(R,T)$ theory turns out to be a bit larger than the value of the Hubble parameter in $\kappa(R,T)$ for a similar functional dependence on $T$. Nevertheless, the exact discrepancy with respect to the GR case is not clear since the authors do not constrain the value of $\lambda$.\\ 
On the other hand, to see how the density evolves with time for a dust universe in the $\kappa(R,T)$ theory, we come back to the fluid equation (\ref{Fluid}) which for dust ($w=0$) reduce to the expression
\begin{equation}\label{Dust}
\dot{\rho}+\frac{2\rho(1-\frac{2\rho}{3\rho_{\max}})}{t(1-\frac{\rho}{3\rho_{\max}})}=0
\end{equation}
Where we have employed the identity $\rho_{m}=3\rho_{\max}$. To solve this equation, first notice that
\begin{equation}
 \frac{1}{2}\leq\frac{1-\frac{2\rho}{3\rho_{\max}}}{1-\frac{\rho}{3\rho_{\max}}}\leq 1
\end{equation}
Where the minimum value $1/2$ of this function is reached for $\rho=\rho_{\max}$ and the maximum value is approached for $\rho<<\rho_{\max}$. Then, an acceptable solution can be provided if we set, $\frac{1-\frac{2\rho}{3\rho_{\max}}}{1-\frac{\rho}{3\rho_{\max}}}\simeq N$ where $N$ is a number between $1/2\leq N\leq 1$. Assuming that the density obeys a relation of the type $\rho=Ct^{\alpha}$,  where $C$ is a constant, and substituting this in Eq.(\ref{Dust}), we obtain the condition $\alpha+2N=0$, which implies
\begin{equation}
\rho(t)\simeq\frac{C}{t^{2N}} \quad 1/2\leq N\leq 1  
\end{equation}
At low densities $N\simeq 1$, and for such a regime the density evolves with time as $\rho\simeq C t^{-2}$, like the standard dust solution of GR.
\subsubsection{\textbf{Stationary solutions and exponential expansion}}
By setting $H=H_{0}=const.$ into Eq. (\ref{H parameter}) and rearranging terms we find an algebraic quadratic equation for the density given by 
\begin{equation}
\rho^{2}-\frac{8\pi G}{\lambda(1-3w)}\rho+\frac{1}{\lambda(1-3w)}\Big(3H_{0}^{2}-\Lambda\Big)=0
\end{equation}
The solutions of this equation are
\begin{equation}
\rho=\frac{\rho_{\max}}{2}\pm\frac{1}{2}\sqrt{\rho^{2}_{\max}-\frac{4}{\lambda(1-3w)}\Big(3H_{0}^{2}-\Lambda\Big)}
\end{equation}
Where $\rho_{\max}$ is the maximum density deduced in the previous subsection. Therefore, for $H_{0}=\sqrt{\Lambda/3}$ we have $\rho=0$ which is the standard vacuum de Sitter solution, and $\rho=\rho_{\max}$. Moreover, the difference, $3H_{0}^{2}-\Lambda$ is bounded from above. Indeed, we can rewrite the last equation in the form
\begin{equation}
\rho=\frac{\rho_{\max}}{2}\pm\frac{1}{2}\sqrt{\rho_{\max}\Big(\rho_{\max}-\frac{1}{2\pi G}\Big(3H_{0}^{2}-\Lambda\Big)\Big)}
\end{equation}
Therefore,
\begin{equation}
3H_{0}^{2}-\Lambda\leq2\pi G\rho_{\max}
\end{equation}
Another interesting feature of this model is the prediction of a specific value of the density (depending on $\lambda$) for the exponential expansion governed by the cosmological constant, which will be exactly equal to $\rho_{\max}$. Indeed, the acceleration equation (\ref{AccelerationEquation}) reduces to $\ddot{a}/{a}=\Lambda/3$ (which implies $a(t)\sim \exp(\sqrt{\Lambda /3} t)$) for $\rho_{\inf}=8\pi G/\lambda(1-3w)=\rho_{\max}$. Then, the inflation takes place when the density reaches the maximum value.  
\subsection{Matter-curvature coupling}
Here we study cosmological solutions for the theory $\kappa(R)=8\pi G+\alpha R$, where $R$ is the curvature scalar and $\alpha$ a constant with units of the inverse of the density. The field equations with such a choice for $\kappa(R,T)$ are
\begin{equation}
R_{\mu\nu}-\frac{1}{2}Rg_{\mu\nu}-\Lambda g_{\mu\nu}=\Big(8\pi G+\alpha R\Big)T_{\mu\nu}
\end{equation}
With a bit of algebra, we find that the expansion rate equation for this model is given by
\begin{equation}\label{Expansionrate}
H^{2}=\frac{8\pi G}{3}\rho\Big(1-f(\rho)(1-3w)\Big)+\frac{\Lambda}{3}\Big(1-4f(\rho)\Big)-\frac{1}{K^{2}a^{2}}
\end{equation}
Where $H\equiv\dot{a}/a$, and $f(\rho)$ is a function of the density given by
\begin{equation}
f(\rho)=\frac{\rho}{\rho_{0}}\Big(\frac{1}{1+(1-3w)\frac{\rho}{\rho_{0}}}\Big)
\end{equation} 
The constant $\rho_{0}$ is a certain density denoted by, $\rho_{0}=1/\alpha$. Since we expect $\alpha$ to be very small, $\rho_{0}$ will presumably take a very large value. For $\rho<<\rho_{0}$, $f(\rho)\approx 0$ and we recover the result of GR for the expansion rate equation. At $\rho=\rho_{0}$ , the function $f(\rho)$ is always regular for all the physically acceptable values of $w$, and $H^{2}$ is also regular as required. Indeed, we have that $f(\rho_{0})=1/(2-3w)$ which is singular only if $w=2/3$. The expansion rate parameter at $\rho=\rho_{0}$ becomes 
\begin{equation}
H^{2}=8\pi G\rho_{0}\Big(\frac{1-2w}{2-3w}\Big)+\frac{\Lambda}{3}\Big(\frac{3w+2}{3w-2}\Big)-\frac{1}{K^{2}a^{2}}
\end{equation}
It is important to note that this model also implies that in general there should exist bounds for the density, namely, the density cannot take an infinite value (with the exception of some particular case). Indeed, for $\rho>>\rho_{0}$ we have that $f(\rho)\approx 1/(1-3w)$. Therefore the first contribution due to the density in Eq.(\ref{Expansionrate}) vanishes identically, and the Hubble parameter becomes for such a regime
\begin{equation}
 H^{2}\simeq \Lambda\Big(\frac{w+1}{3w-1}\Big)-\frac{1}{K^{2}a^{2}}
\end{equation}
which is always negative for $w<1/3$. (The special case $w=1/3$ will be analyzed apart). Therefore, the limit $\rho>>\rho_{0}$ lacks physical sense, and we conclude that the density cannot be much larger than $\rho_{0}=1/\alpha$ when $w<1/3$.\\
On the other hand, the case $w=1/3$ corresponds to a radiation-dominated universe, the function $f$ for such a value of the parameter $w$ becomes, $f(\rho)=\rho/\rho_{0}$. Taking this into account and setting $w=1/3$ into Eq.(\ref{Expansionrate}), we have
\begin{equation}
 H^{2}=\frac{4}{3}\rho\Big(2\pi G-\frac{\Lambda}{\rho_{0}}\Big)+\frac{\Lambda}{3}-\frac{1}{K^{2}a^{2}}
\end{equation}
In this case, there are no bounds for the growth of the density because the assumed smallness of $\Lambda/\rho_{0}$ compared to $2\pi G$ assures that the quadratic density term will not acquire a negative sign.\\
Regarding the generalized Friedmann acceleration equation for this theory, we have
\begin{equation}
\frac{\ddot{a}}{a}\Big(1-\frac{\rho}{\rho_{0}}(3w+1)\Big)=\frac{\Lambda}{3}+\rho(3w+1)\Big(-\frac{4\pi G}{3}+\frac{1}{\rho_{0}}\Big(H^{2}+\frac{1}{K^{2}a^{2}}\Big)\Big)
\end{equation}
Where the scale factor $H^{2}$ is provided by Eq.(\ref{Expansionrate}). The acceleration equation is difficult to interpret given the quantity of terms involved. For $\rho<<\rho_{0}$ we obviously recover the GR result, namely, the cosmological constant rules the late-time cosmic speed-up. It is worth noting that the term that accounts for the accelerated expansion vanishes when $\rho=\rho_{0}/(3w+1)$. It turns out that for such a specific value of the density the effects of the accelerated expansion are null.\\
Let us conclude with some brief comments regarding the theory $\kappa(R)=8\pi G-\alpha R$. By solving again the equations for the FLRW metric and the perfect fluid, we find a similar expression for the modified expansion rate equation which is explicitly given by
\begin{equation}
H^{2}=\frac{8\pi G}{3}\rho\Big(1+g(\rho)(1-3w)\Big)+\frac{\Lambda}{3}\Big(1+4g(\rho)\Big)-\frac{1}{K^{2}a^{2}}  
\end{equation}
Where the function $g(\rho)$ is denoted by
\begin{equation}
g(\rho)=\frac{\rho}{\rho_{0}}\Big(\frac{1}{1-(1-3w)\frac{\rho}{\rho_{0}}}\Big)
\end{equation}
The difference with respect to the theory $\kappa(T)=\kappa+\alpha R$ discussed before, lies in the fact that the latter is better in terms of singularities. In particular, the main difference is due to the behavior of the functions $f(\rho)$ and $g(\rho)$ regarding regular properties. In contrast to $f(\rho)$, that was regular everywhere and in particular at $\rho=\rho_{0}/(1-3w)$, the function $g(\rho)$ is singular at such points and therefore $H^{2}$ for the model $\kappa(R)=\kappa-\alpha R$ will also diverge for that family of values of the density. Talking about the divergences of $f(\rho)$ and $g(\rho)$ is equivalent to talk about the divergences of the curvature scalar $R$ for these models (see the appendix for details).
\subsection{Static spherically symmetric perfect fluid}
Spherically symmetric scenarios in GR and their extensions are very important because a wide range of phenomena such as black holes, neutron/quark stars and gravitational collapse can be theoretically described by means of spherical symmetry. The aim of this subsection is to present the main equations for the model $\kappa(T)=8\pi G-\lambda T$, and to show how the a unknown metric components can be expressed in terms of the modified sources. Nonetheless, the explicit solution of the equations is an open problem that we leave for future works. The line element for a static spherically symmetric space-time takes the form
\begin{equation}
ds^{2}=A^{2}(r)dt^{2}-B^{-1}(r)dr^{2}-r^{2}\Big(d\theta^{2}+\sin^{2}\theta d\varphi^{2}\Big)
\end{equation}
By solving the field equations of the theory (\ref{FieldEquations2}) for $\Lambda=0$, we obtain the system
\begin{equation}
\frac{1-B}{r^{2}}-\frac{1}{r}\frac{dB}{dr}=8\pi G\rho\Big(1-\frac{1}{\rho_{m}}(\rho-3p)\Big)
\end{equation}
\begin{equation}
\frac{2}{A}\frac{dA}{dr}\frac{B}{r}+\frac{B-1}{r^{2}}=8\pi G p\Big(1-\frac{1}{\rho_{m}}(\rho-3p)\Big)
\end{equation}
Where we already denoted the constant $\rho_{m}$ as $\rho_{m}\equiv8\pi G/\lambda$.
The non-conservation of the stress-energy tensor implies another first order differential equation given by
\begin{equation}
p^{\prime}+\frac{A^{\prime}}{A}\Big(p+\rho\Big)=\frac{\rho^{\prime}-3p^{\prime}}{3+\frac{\rho_{m}-\rho}{p}}
\end{equation}
Where $p^{\prime}\equiv dp/dr$. The presence of a non-null right hand side represents the departure with respect to the GR case. On the other hand, the integration of the first equation gives
\begin{equation}
B(r)=1-\frac{2Gm_{eff}(r)}{r}
\end{equation}
Where
\begin{equation} 
m_{eff}(r)=4\pi\int_{0}^{r}\rho(z)\Big(1-\frac{1}{\rho_{m}}(\rho(z)-3p(z))\Big)z^{2}dz
\end{equation}
Represents the ``effective mass'' content of the distribution within the sphere of radius $r$. The remaining unknown metric component, the function $A(r)$, can also be expressed as an integral over the matter content and their density and pressure. Indeed, combining the first and second equations, we obtain
\begin{equation}
8\pi G\Big(\rho+p\Big)\Big(1-\frac{1}{\rho_{m}}(\rho-3p)\Big)=\frac{2}{A}\frac{dA}{dr}\frac{B}{r}-\frac{1}{r}\frac{dB}{dr}
\end{equation}
This equation can be immediately integrated to give
\begin{equation}
A(r)=C\sqrt{B(r)}\exp\Big(4\pi G \int \frac{(\rho+p)\Big(1-\frac{1}{\rho_{m}}(\rho-3p)\Big)}{B}rdr\Big)
\end{equation}
With $C$ an arbitrary integration constant. Therefore, the unknown metric components have been expressed in terms of the modified sources, as required. It is important to recall that the internal Schwarzschild-type solution should match with the external one at $r=R$, being $R$ the radius of the compact object. Since the external Schwarzschild solution satisfies $A(r)=\sqrt{B(r)}$, and $B(r)=1-2GM/r$, we see that the obtained internal metric components have the appropriate structure to match with the external solution, and this fixes the value of the constant $C$ to be
\begin{equation}
C^{-1}=\exp\Big(4\pi G \int \frac{(\rho+p)\Big(1-\frac{1}{\rho_{m}}(\rho-3p)\Big)}{B}rdr\Big)\vert_{r=R} 
\end{equation}
\subsection{Comparison among $\kappa(R,T)$ and $f(R,T)$ theories}
It would be interesting to compare the theory $\kappa(R,T)$ to a gravitational Lagrangian theory, for example, we can compare the special case $\kappa(R,T)=\kappa(T)$ to a subclass of the variational theories $f(R,T)$, which are also non-conservative theories (in the sense that $\nabla_{\nu}T^{\mu\nu}\neq 0$). One of the most natural choices is a model of the type $f(R,T)=R+f_{2}(T)$. For a perfect fluid, this model leads to the field equations\cite{Harko}
\begin{equation}
 R_{\mu\nu}-\frac{1}{2}Rg_{\mu\nu}=\Big(8\pi G+f_{2}^{\prime}(T)\Big)T_{\mu\nu}+T_{\mu\nu}^{eff}
\end{equation}
Where $T_{\mu\nu}^{eff}$ for this particular theory $f(R,T)=R+f_{2}(T)$ is given by
\begin{equation}
T_{\mu\nu}^{eff}=\Big(f_{2}^{\prime}(T)p+f_{2}(T)\Big)g_{\mu\nu}
\end{equation}
Then, the comparison to the field equations
\begin{equation}
R_{\mu\nu}-\frac{1}{2}Rg_{\mu\nu}-\Lambda g_{\mu\nu}=\kappa(T)T_{\mu\nu}
\end{equation}
leads to the conclusion that the field equations of both theories only match if we assume in the field equations of the $\kappa(T)$ theory\footnote{Note that if we consider the variable cosmological term in the action we will have a different theory} the possibility of a variable cosmological term $\Lambda(T)$ that depends on the matter sources, namely
\begin{equation}
\Lambda(T)\equiv f_{2}^{\prime}(T)p+f_{2}(T)
\end{equation}
It has been pointed out that recent cosmological data favor a variable cosmological constant\cite{Poplawski}.
\subsubsection{\textbf{Modified geodesic equation of motion}}
The fact that the stress-energy tensor is not covariantly conserved has the effect of modifying the equations of motion of particles. In fact, as a direct consequence an extra force will arise in the geodesic equation. To see in detail how this happens, we can write rewrite Eq.(\ref{nonconservation2}) for the perfect fluid in the form

\begin{equation}
\small
u^{\mu}u^{\nu}\nabla_{\nu}(p+\rho)+(p+\rho)\Big(u^{\mu}\nabla_{\nu}u^{\nu}+u^{\nu}\nabla_{\nu}u^{\mu}\Big)-g^{\mu\nu}\nabla_{u}p=-\frac{\nabla_{\nu}\kappa}{\kappa}T^{\mu\nu}
\end{equation}
Where $\nabla_{\nu}g^{\mu\nu}=0.$ Let us introduce now an auxiliary metric $h_{\mu\lambda}$ defined by $h_{\mu\lambda}\equiv g_{\mu\lambda}-u_{\mu}u_{\lambda}$. Then, multiplying the last equation by $h_{\mu\lambda}$ we have
\begin{equation}\label{modified}
g_{\mu\lambda}u^{\nu}\nabla_{\nu}u^{\mu}=\frac{\nabla_{\nu}[\kappa(R,T)p]}{(p+\rho)\kappa(R,T)}h^{\nu}_{\lambda}
\end{equation}
Where we have used the identities, $h_{\mu\lambda}T^{\mu\nu}=-h^{\nu}_{\lambda}p$, and $h_{\mu\lambda}u^{\mu}=0$.
Therefore, with the aid of the identity
\begin{equation}
u^{\nu}\nabla_{\nu}u^{\mu}=\frac{d^{2} x^{\mu}}{ds^{2}}+\Gamma^{\mu}_{\nu\lambda}u^{\nu}u^{\lambda}
\end{equation}
where $\Gamma^{\mu}_{\nu\lambda}$ is the Levi-Civita connection of $g^{\mu\nu}$, the modified geodesic equation of motion (\ref{modified}) acquires the form
\begin{equation}
\frac{d^{2} x^{\mu}}{ds^{2}}+\Gamma^{\mu}_{\nu\lambda}u^{\nu}u^{\lambda}=f^{\mu}
\end{equation}
Where $f^{\mu}$ denotes a four-vector ``force'' given by
\begin{equation}\label{extraforce}
f^{\mu}=\frac{\nabla_{\nu}[\kappa(R,T)p]}{(p+\rho)\kappa(R,T)}\Big(g^{\mu\nu}-u^{\mu}u^{\nu}\Big)
\end{equation}
Therefore, for dust ($p\approx0$), we recover the geodesic equation of GR. Moreover, when $\kappa(R,T)= const=8\pi G$, the standard result of GR for perfect fluids with pressure is recovered as well. Notice that the vector $f^{\mu}$ is orthogonal to $u^{\mu}$, namely, $f^{\mu}u_{\mu}=0$. Eq.(\ref{extraforce}) suggests to re-define the pressure and density as, $\rho_{eff}=\kappa(R,T)\rho$, and $p_{eff}=\kappa(R,T)p$. Doing this, the form of the geodesic equation is identical to the GR case with the new variables $\rho_{eff}$ and $p_{eff}$ playing the role of $p$ and $\rho$. On the other hand, it is worth comparing this extra force with other external force that arises in the geodesic equation of motion of the Lagrangian theory $f(R,T)$. For this theory, the extra force is given by\cite{Harko}:
\begin{equation}
f^{\mu}_{f(R,T)}=8\pi G\frac{\nabla_{u}p}{(p+\rho)[8\pi G+f_{T}(R,T)]}\Big(g^{\mu\nu}-u^{\mu}u^{\nu}\Big)
\end{equation}
Where $f_{T}(R,T)=\partial f(R,T)/\partial T$. In order to see explicitly the differences among the extra force in both theories, we can select the same dependence on the trace $T$ to compare predictions. For example, setting $\kappa(R,T)=\kappa(T)=8\pi G-\lambda T$, and $f(R,T)=f_{1}(R)-\beta T$, where $\lambda$, $\beta$ are constants and $f_{1}(R)$ is an arbitrary function of $R$. Then, a first important consequence is that for $T=0$ (photons), both theories predict the same extra force, but when $T\neq 0$ such is the case of massive particles, the two forces are different. Therefore, a detailed investigation on the trajectory of massive particles in a gravitational field could help to find out which among these different theories represent the most viable generalization of Einstein's GR. 
\subsection{Generalized energy conditions}
Generalized energy conditions (GEC) in Extended Theories of Gravity have been studied in detail in several works, see for example \cite{Cappo,Alva,Santos,ecsetare}. To study the role of the energy conditions in the $\kappa(R,T)$ theory, with the aim to investigate if they are violated or not, it is convenient to recast the field equations in the form:
\begin{equation}
R_{\mu\nu}-\frac{1}{2}R g_{\mu\nu}=T_{\mu\nu}^{eff}
\end{equation}
Where the effective energy momentum tensor $T_{\mu\nu}^{eff}$ is defined by
\begin{equation}
T_{\mu\nu}^{eff}=\kappa(R,T)T_{\mu\nu}+\Lambda g_{\mu\nu}
\end{equation}
The effective energy momentum tensor $T_{\mu\nu}^{eff}$, in turn allows to define the effective pressure $p_{eff}$ and density $\rho_{eff}$ necessary to present the conditions required for realizing each type of the energy conditions. In fact, $p_{eff}$ and $\rho_{eff}$ will be very similar to those that arose in the previous subsection. Indeed, by assuming that the content of the universe behaves like a perfect fluid, and for a flat FRLW metric we have
\begin{equation}
3H^{2}=\kappa(R,T)\rho+\Lambda
\end{equation}
\begin{equation}
-2\dot{H}-3H^{2}=\kappa(R,T)p-\Lambda
\end{equation}
It is convenient to focus on the case $\kappa(R,T)=\kappa(T)$ in order to simplify the analysis. This choice for $\kappa(R,T)$ allows one to define the effective pressure and density as
$\rho_{eff}=\kappa(T)\rho+\Lambda$, $p_{eff}=\kappa(T)p-\Lambda$. By using these expressions for $\rho_{eff}$ and $p_{eff}$, we get the null energy condition (NEC), the weak energy condition (WEC), the strong energy condition (SNC) and the dominant energy condition (DEC) \cite{Alva} as:
\small
\begin{eqnarray}
\mbox{NEC:}\quad\quad\quad\quad\rho_{eff}+p_{eff}\geq 0 \,\,\,,\label{vincent14}
\end{eqnarray}
\begin{eqnarray}
\mbox{WEC:}\quad\quad\quad\quad\quad\quad \rho_{eff}\geq 0\,\,,\quad \rho_{eff}+p_{eff}\geq 0\,\,\,,\label{vincent15}
\end{eqnarray}
\begin{eqnarray}
\mbox{SEC:}\quad\quad\quad\rho_{eff}+3p_{eff}\geq 0 \,\,,\quad \rho_{eff}+p_{eff}\geq 0\,\,,\label{vincent16}
\end{eqnarray}
\begin{eqnarray}
\mbox{DEC:}\quad\rho_{eff}-p_{eff}\geq 0\,\,,\quad \rho_{eff}+p_{eff}\geq 0\,\,,\quad \rho_{eff}\geq 0\,\,.\label{vincent17}
\end{eqnarray}
\normalsize
Then, for a general $\kappa(T)$ model, the GEC will acquire the explicit expressions:
\small
\begin{eqnarray}
\mbox{NEC:}\quad\quad\quad\quad\quad\kappa(T)(\rho+p)\geq 0 \,\,\,,\label{vincent14}
\end{eqnarray}
\begin{eqnarray}
\mbox{WEC:}\quad\quad\quad\quad\kappa(T)\rho+\Lambda\geq 0\,\,,\quad \rho_{eff}+p_{eff}\geq 0\,\,\,,\label{vincent15}
\end{eqnarray}
\begin{eqnarray}
\mbox{SEC:}\quad\kappa(T)(\rho+3p)-2\Lambda\geq 0 \,\,,\quad \rho_{eff}+p_{eff}\geq 0\,\,,\label{vincent16}
\end{eqnarray}
\begin{eqnarray}
\mbox{DEC:}\quad\kappa(T)(\rho-p)+2\Lambda\geq 0\,\,,\quad \rho_{eff}+p_{eff}\geq 0\,\,,\quad \rho_{eff}\geq 0\,\,.\label{vincent17}
\end{eqnarray}
\normalsize
It is not difficult to prove that for the perfect fluid and the model studied in this work, namely $\kappa(T)=8\pi G-\lambda T$, the GEC are satisfied, since the density is bounded. Indeed, the NEC will be
\begin{equation}
\mbox{NEC:}\quad \Big(8\pi G-\lambda(1-3w)\rho\Big)\rho(1+w)\geq 0
\end{equation}
Where we have neglected the contribution of the cosmological constant. For $w>-1$, the NEC is automatically fulfilled if $\rho\leq \rho_{\max}$, where $\rho_{\max}=8\pi G/\lambda(1-3w)$ in agreement with Eq.(\ref{maximum density}). Regarding the WEK, it is also satisfied identically if $\rho\leq \rho_{\max}$. As for the SEC, we have:
\small
\begin{align}
\mbox{SEC:}\Big(8\pi G-\lambda(1-3w)\rho\Big)\rho(1+3w)\geq 0\,\nonumber\\
\quad\Big(8\pi G-\lambda(1-3w)\rho\Big)\rho(1+w)\geq 0 
\end{align}
\normalsize
For $w > -1/3$, $\rho\leq \rho_{\max}$ the SEC is realized as well. Finally, the DEC acquires the form
\small
\begin{equation}
\mbox{DEC:}\Big(8\pi G-\lambda(1-3w)\rho\Big)\rho(1-w)\geq 0\,\,,\quad \rho_{eff}+p_{eff}\geq 0\,\,,\quad \rho_{eff}\geq 0\,\,.\label{vincent17}
\end{equation}
\normalsize
For $w<1$ the DEC is also satisfied for $\rho\leq \rho_{\max}$. Therefore, the presence of a bound for the density guarantees that all the energy conditions are satisfied in a consistent way for the model $\kappa(T)=8\pi G-\lambda T$.
\section{SUMMARY AND CONCLUSIONS}
The important degree of arbitrariness inherent in the choice of the gravity Lagrangian has lead to a large amount of different modified gravity proposals, many of which are so similar that it is difficult to distinguish one from the other. The Lagrangian formalism has undoubted advantages at the level of symmetries implementation and conservation-laws derivation, but possible theoretical alternatives to standard Lagrangian theories also deserve consideration. In this sense, the importance of Non-Lagrangian theories in other branches of theoretical physics such as quantum field theory is being acknowledged in the last years. Among their advantages, it seems increasingly clear that these theories offer new opportunities in the search of new types of invariants. \\
In this work, and in absence of a foundational principle, we have explored an example of a Non-Lagrangian modified gravity theory inspired by Maxwell's approach to Electrodynamics, adding new possible source terms directly in the field equations, namely, we have investigated a gravitational analogue of the Maxwellian ``displacement current'' contribution. It should be noted that our approach does not mean that a variational formulation of the theory could not exist, but in this work we did not focus on that problem. In particular, we have analyzed some special cases that belong to the classification: $\kappa(R,T)=k+f(T)$ and $\kappa(R,T)=k+f(R)$, which corresponds to matter-matter and matter-curvature couplings respectively. We carried out a preliminary study of some cosmological aspects of these models in a FLRW universe filled by a perfect fluid, and it was shown that the density in bounded from above in some of them. Furthermore, the formal similarities and differences among the theory $\kappa(R,T)$ and the Lagrangian theory $f(R,T)$ were also investigated. The field equations can match in some particular cases that imply a variable cosmological term that depends on the energy-matter content. However, both theories are essentially different at the level of the equations of motion for massive particles. Moreover, the generalized energy conditions were also investigated for the theory $\kappa(T)=8\pi G-\lambda T$ and we have shown that the existence of a maximum density $\rho_{\max}$ guarantees that all the energy conditions are satisfied in a consistent way. In summary, we have presented in this work an example of a Non-Lagrangian modified gravity theory, which is a relatively unexplored research avenue in the field of modified gravity.
\begin{acknowledgments}
It is a pleasure to thank Prof. S. Odintsov for pointing me out Ref.\cite{Mafia} and Prof. H. Shabani for bringing to my attention Refs.\cite{Shabani,Shabani2,Shabani3}. I thank Prof. G. J. Olmo and Prof. F. Rahaman for stimulating discussions and encouragement. I also thank the anonymous referees for comments that helped me to improve the quality of the manuscript. 
\end{acknowledgments}
\section{APPENDIX. DIVERGENCES AND ZEROS OF THE CURVATURE SCALAR}
By contracting the field equations (\ref{FieldEquations}) with $g^{\mu\nu}$ it is easy to obtain a generic relation among the traces given by
\begin{equation}
-R-4\Lambda=\kappa(R,T)T
\end{equation}
Then, the exact relation among $R$ and $T$ requires that we fix a particular $\kappa(R,T)$ model. Choosing a model of the type $\kappa(T)=8\pi G-\lambda T$, we find that there exists an algebraic quadratic equation among $R$ and $T$ given by
\begin{equation}
R=\lambda T^{2}-8\pi G T-4\Lambda
\end{equation}
In the limit $\Lambda\rightarrow 0$ we obtain
\begin{equation}
R\simeq \lambda T\Big(T-\frac{8\pi G}{\lambda}\Big)
\end{equation}
Therefore $R$ vanishes for $T=0$ (the vacuum solution as in GR), and for $T=8\pi G/\lambda$. In general, $R$ vanishes for specific values of the pressure and the density that are solutions of the quadratic equation, $\lambda T^{2}-8\pi G T-4\Lambda=0$. Given that $\rho_{m}\equiv 8\pi G/\lambda$, $T=\rho-3p$ for the perfect fluid, and using a barotropic equation of state of the type $p=w\rho$, we obtain such specific values of the density where $R=0$
\begin{align}
\rho=&\displaystyle\frac{1}{2(1-3w)}\Big(\rho_{m}\pm\sqrt{\rho_{m}^{2}+\frac{16\Lambda}{\lambda}}\Big)\nonumber\\
&=\displaystyle\frac{1}{2(1-3w)}\rho_{m}\Big(1\pm\sqrt{1+\frac{16\Lambda}{\lambda \rho_{m}^{2}}}\Big)
\end{align} 
For $w\neq 1/3$. If $\rho_{m}>>4\sqrt{\Lambda/\lambda}$, we can approximate the solutions as
\begin{equation}
\rho_{1}\simeq\frac{1}{1-3w}\Big(\rho_{m}+\frac{4\Lambda}{\lambda\rho_{m}}\Big)
\end{equation}º
\begin{equation}
\rho_{2}\simeq -\frac{4\Lambda}{(1-3w)\lambda\rho_{m}}
\end{equation}
Therefore, we see that the values of the density that vanish $R$ are purely mathematical an not physical solutions. Recall that $\rho$ is bounded and verifies, $\rho\leq \rho_{\max}$, namely, $\rho\leq\rho_{m}/(1-3w)$ according to Eq.(\ref{bounds}) In the limit $\Lambda/\lambda\rightarrow 0$, we obtain that $\rho_{2}=0$ (as in GR), and $\rho_{1}=\rho_{m}/(1-3w)=\rho_{\max}$, which is the extra solution with respect to the GR case. 
On the other hand, for the pair of models $\kappa(R)=k\pm \alpha R$, the exact relation among the traces is
\begin{equation}
R=\frac{-4\Lambda-8\pi G T}{1\pm \alpha T}
\end{equation}
This relation becomes, with the assumed approximations,
\begin{equation}
R=\frac{-4\Lambda-8\pi G \rho(1-3w)}{1\pm(1-3w)\frac{\rho}{\rho_{0}}}
\end{equation}
Where, $\rho_{0}\equiv 1/\alpha$. Then, $R$ has a zero at a density given by $\rho=\Lambda/2\pi G(3w-1)$ which is negative for $w<1/3$. Regarding the divergences, the curvature scalar is regular everywhere for the theory $\kappa(R)=k+\alpha R$ (with the exception of the special case $w=2/3$, which contains a pole at $\rho=\rho_{0}$). On the contrary, the model $\kappa(R)=k-\alpha R$ yields to divergences for the family of values of the density given by $\rho=\rho_{0}/(1-3w)$


\begin{thebibliography}{99}
\bibitem{Sauer}Sauer, T. "Albert Einstein's 1916 Review Article on General Relativity". arXiv:physics/0405066
\bibitem{Renn} Renn,J. and Schemmel,M. "The Genesis of General Relativity", (Springer, Berlin, 2007) 
\bibitem{Einstein}Einstein, A. "Die Feldgleichungen der Gravitation". Sitzungsberichte der Preussischen Akademie der Wissenschaften zu Berlin, S.844–847, 25. November 1915
\bibitem{Hilbert} Hilbert, D. (1915) "Die Grundlagen der Physik". Konigl. Gesell. d. Wiss. Göttingen, Nachr. Math.-Phys. Kl. 395-407
\bibitem{Maxwell} Maxwell, J. C. (1861). "On physical lines of force". Philosophical Magazine. 90: 11–23 
\bibitem{Buchdahl} Buchdahl, H. A. (1970). "Non-linear Lagrangians and cosmological theory". Monthly Notices of the Royal Astronomical Society. 150: 1–8
\bibitem{DeFelice}De Felice, A; Tsujikawa, S. (2010). "f(R) Theories". Living Reviews in Relativity. 13. arXiv:1002.4928
Applied Maths. and Compt. 112, No. 1 (2000) 63-73.
\bibitem{Capo} S. Capozziello and M. De Laurentis."Extended Theories
of Gravity". Phys.Rept.509, 167(2011).arXiv:1108.6266
\bibitem{Olmo}Olmo,G.J."Palatini Approach to modified Gravity.f(R)Theories and Beyond". Int.J.Mod.Phys.D 20, 413–462(2011).
\bibitem{Lobo}Lobo,F.S.N. "The Dark side of gravity: Modified theories
of gravity", arXiv:0807.1640 [gr-qc].
\bibitem{Sotiriou}Sotiriou,P; Faraoini,V. "f(R) Theories Of Gravity". arXiv:0805.1726
\bibitem{Mafia}Odintsov,S; Nojiri,S. "Unified cosmic history in modified gravity: from F(R) theory to Lorentz non-invariant models" Phys.Rept.505:59-144,2011. arXiv:1011.0544
\bibitem{Rastall}Rastall,P. "Generalization of the Einstein theory", Physical Review D6 (1972) 3357-3359
\bibitem{Fabris}Fabris, J.C; Piatella, O.F; Rodrigues, D.C; Batista, C.E.M; Daouda, M.H; "Rastall Cosmology"  Int. J. Mod. Phys. Conf. Ser. 18, 67 (2012)
\bibitem{Harko}Harko,T; Lobo,F.S.N; Nojiri,S; Odintsov,S.D. "f(R,T) gravity" arXiv:1104.2669
\bibitem{Gao}hangjun Gao, Fengquan Wu, Xuelei Chen, and You-Gen Shen
Phys. Rev. D 79, 043511(2009)
\bibitem{Nojiri2}Nojiri, S; Odintsov, S.D. Eur. Phys. J. C (2017) 77: 528. https://doi.org/10.1140/epjc/s10052-017-5097-x
\bibitem{Saridakis}Ricci-Gauss-Bonnet holographic dark energy, E. N. Saridakis
Phys. Rev. D 97, 064035 (2018)
\bibitem{Sharif}Sharif,M; Zubair,M. "Thermodynamics in f(R,T) theory of gravity" arXiv:1204.0848
\bibitem{Shabani}Shabani, H; Farhoudi, M. "$f(R,T)$ Cosmological models in phase space" Phys. Rev. D 88, 044048
\bibitem{Shabani2}Shabani,H; Farhoudi,M. "Cosmological and solar system consequences of $f(R,T)$ gravity models" Phys. Rev. D 90, 044031
\bibitem{Shabani3}Shabani,H; Hadi Ziae, A. "Late-time cosmological evolution of a general class of $f(R,T)$ gravity with minimal curvature-matter coupling".Eur. Phys. J. C 77:507 (2017) arXiv:1703.06522 
\bibitem{Gukov} Gukov, S. J. High Energ. Phys. (2017) 2017: 178.
\bibitem{Razamat}Abhijit Gadde, Shlomo S. Razamat, and Brian Willett, " "Lagrangian" for a Non-Lagrangian Field Theory with $N=2$ Supersymmetry". Phys. Rev. Lett. 115, (2015) 
\bibitem{Brans} Brans, C. H.; Dicke, R. H. (1961). "Mach's Principle and a Relativistic Theory of Gravitation". Physical Review. 124 (3): 925–935.
\bibitem{Brans2}Brans, C.H; "The roots of scalar-tensor theory: an approximate history." arXiv:gr-qc/0506063
\bibitem{Moffat}Moffat, J.W; "Scalar-Tensor-Vector Gravity Theory." arXiv:gr-qc/0506021
\bibitem{Poplawski}Poplawski,N.J. arXiv:gr-qc/0608031
\bibitem{Olmo2}Beltran-Gimenez, J; Heisenberg, L; Olmo,G.J; Rubiera-Garcia, D. "Born-Infeld inspired modifications of gravity" arXiv:1704.03351
\bibitem{Alva}F.Alvarenga, M.Houndjo, A.Monwanou and J.Orou, "Testing Some f(R,T) Gravity Models from Energy Conditions," Journal of Modern Physics, Vol. 4 No. 1, 2013, pp. 130-139.
\bibitem{Cappo}S.Capozziello, F. S.?N. Lobo, J. P. Mimoso, "Generalized energy conditions in extended theories of gravity" Phys. Rev. D 91, 124019 (2015)
\bibitem{Santos}  J.
Santos, J.S. Alcaniz, N. Pires and M.J. Reboucas, Phys.
Rev. D 75, (2007) 083523; A. A. Sen and R. J. Scherrer,
Phys. Lett. B 659, 457 (2008); J. Santos, J. S. Alcaniz,
J.H. Kung, Phys. Rev. D 52, (1995) 6922; Phys. Rev.
D 53, (1996) 3017
\bibitem{ecsetare} N. M. Garc\'{i}a, T. Harko, F. S. N. Lobo and J. P. Mimoso, ``  Energy conditions in modified Gauss-Bonnet gravity", [arXiv: 1011.4159v2[gr-qc]]. 
A. Banijamali, B. Fazlpour and M. R. Setare, ``Energy conditions in f(G) modified gravity with non-minimal coupling to matter", Astrophys. Space Sci. 338 (2012) 327-332. arXiv: 1111.3878 [physics.gen-ph].
\end{thebibliography}
\end{document}